\documentstyle[psfig,subfigure]{mn}

\begin{document}
\title[XMM-Newton Surveys of the CFRS Fields - I]{XMM-Newton Surveys
of the CFRS Fields - I: The Sub-mm/X-ray relation
\thanks{Based on observations obtained with {\em XMM-Newton\/},
an ESA science mission with instruments and contributions directly
funded by ESA Member States and NASA}}
\author[Waskett et al.]
{Timothy J. Waskett$^{1}$, 
Stephen A. Eales$^{1}$, 
Walter K. Gear$^{1}$, \cr 
Elizabeth M. Puchnarewicz$^{2}$, 
Simon Lilly$^{3}$, 
Hector Flores$^{4}$, \cr 
Tracy Webb$^{5}$, 
David Clements$^{6}$,
Jason A. Stevens$^{7}$,
Trinh X. Thuan$^{8}$\\
$^1$Department of Physics and Astronomy, University of Wales Cardiff,
PO Box 913, Cardiff, CF24 3YB, UK\\ 
$^2$ Mullard Space Science Laboratory, University College London, UK\\ 
$^3$ Herzberg Institute for Astrophysics, Dominion Astronomical
Observatory, National Research Council, Canada\\ 
$^4$ Observatoire de Paris, Section de Meudon, DAEC, 92195 Meudon
Principal Cedex, France\\ 
$^5$ Department of Astronomy and Astrophysics, University of Toronto, Canada\\ 
$^6$ Physics Department, Blackett Laboratory, Imperial College,
London, UK\\
$^7$ UK Astronomical Technology Centre, Royal Observatory, Blackford
Hill, Edinburgh EH9 3HJ\\
$^8$ Department of Astronomy, University of Virginia, P.O. Box 3818,
Charlottesville, VA 22903}

\maketitle

\begin{abstract}
First results from {\em XMM-Newton\/} observations of the Canada
France Redshift Survey (CFRS) 3hr, 10hr and 14hr fields are presented.
Limited regions of two of the XMM surveys (3 and 14hr) are compared to
the Canada UK Deep sub-mm Surveys (CUDSS) undertaken with SCUBA.  None of
the 27 SCUBA sources in the 3hr field are detected by XMM, while one
of the 23 SCUBA sources in the 14hr field is found to coincide with an 
X-ray source.  The SCUBA population as a whole is not
significantly detected in either the $0.5-2~keV$ or the $2-10~keV$
X-ray bands, even after coadding the X-ray flux at the SCUBA
positions, in both fields.  The 18 X-ray sources within the CUDSS 3hr
map yield a mean sub-mm flux of $0.48\pm0.27~mJy$ after
coadding the sub-mm flux at the X-ray positions.  Using this result we
place an upper limit on the contribution of AGN to the sub-mm
background at $850~\mu$m of $\sim7~per~cent$.  Conversely we estimate
the contribution of sub-mm sources to the $0.5-2~keV$ X-ray background
to be $<16.5~per~cent$.  These results strongly support the conclusion 
that the two backgrounds are caused by different processes, in the
one case nucleosynthesis in stars, in the other accretion onto
black-holes.  We conclude that it is possible for SCUBA
sources in general to contain AGN, as long as they are Compton-thick
and are at $z>2.3$.  The ratio of the X-ray to sub-mm flux for the
X-ray sources however, implies that even when a galaxy does contain an 
AGN, most of the energy heating the dust is from young stars and not
from the active nucleus.
\end{abstract}


\section{Introduction}
Recent advances in sub-mm astronomy have allowed rapid progress in our 
understanding of the early Universe.  This region of the
electromagnetic spectrum has largely been opened up by the powerful
instrumentation that has become available within the last decade.  One 
of the key instruments in this field has been the sub-mm Common User
Bolometer Array (SCUBA), operating primarily at $850~\mu$m, at the
Naysmith focus of the 15m James Clerk Maxwell Telescope (JCMT) on Mauna
Kea in Hawaii.  Since its commission, deep surveys with SCUBA (Smail,
Ivison \& Blain 1997; Hughes et al. 1998; Barger et al. 1998; Eales et 
al. 1999) have resolved a significant fraction of the recently
discovered far-IR and sub-mm background (Puget et al. 1996; Fixsen et
al. 1998; Hauser et al. 1998), also called the Cosmic IR Background
(CIRB).  The importance of this becomes clear when one considers that
the total integrated energy observed in the CIRB is comparable to the
total integrated energy associated with the optical-UV background.
Dust is very important in the interstellar medium as it absorbs
optical-UV photons and re-radiates far-IR photons, thus affecting our
view of the  Universe.  Since this is the mechanism that produces most
of the CIRB, it is possible that half of the light ever emitted by
stars has been reprocessed by dust.  However, there is one other
mechanism that could contribute to this background light and that is
the absorption, by dust, of the radiation from Active Galactic Nuclei 
(AGN).

In the standard model of AGN the central black hole is fed by an
accretion disc (eg. Antonucci 1993).  Around this disc lies a torus of
heavily obscuring material containing large amounts of dust
(eg. Nenkova, Ivezi\'{c} \& Elitzur 2002).  If viewed close to the
axis the AGN may be visible directly, in which case a Quasar or a
Type-I Seyfert Galaxy is observed, with characteristic broad spectroscopic
lines.  Evidence for the obscuring torus comes from observations of
Type-II Seyfert Galaxies, which in this model are seen edge-on.
type-I emission, which would indicate the presence of an AGN, can only
be indirectly observed by reflection off material above and below the
torus.  Direct evidence for the torus itself comes from its
interaction with the nuclear radiation.  The AGN heats up the torus
which produces primarily mid-IR emission, but there will also be
sub-mm emission from the dust.  The process is much the same as in
starlight re-processing, but here it is mainly UV and X-ray photons
that are absorbed, and the dust tends to be hotter.  Models suggest
that between 5 and $30~per~cent$ of the CIRB might be produced in this
way (Gunn \& Shanks 1999; Almaini, Lawrence \& Boyle 1999).
Franceschini, Braito \& Fadda (2002) argue for the revision of this
standard model, and suggest that Type-I and Type-II AGN are different
populations that follow unrelated evolutionary paths, in particular at
high redshifts.  As deeper and more extensive X-ray/mid-IR surveys are
collated this possibility will be tested more thoroughly, and a more
complete picture of AGN may emerge. 

The Cosmic X-ray Background (XRB) on the other hand contains much less
energy than the optical or IR backgrounds ($\sim1/100~th$ depending on 
the definition).  With the advent of powerful new X-ray telescopes
such as {\em Chandra\/} and {\em XMM-Newton\/}, it has become possible 
to resolve most of this background radiation into discrete sources
(eg. Rosati et al. 2002; Hasinger et al. 2001).  The simplest form of
the XRB is a power law defined as,

\[
N = K E^{-\Gamma}
\]
where $N$ is the number of photons per second per $cm^{2}$ per $keV$, $K$ is a
normalisation constant and $\Gamma$ is the photon index.  This reaches
a peak in energy density at $\sim30~keV$, and below this it has a very hard
spectrum with a photon index $\Gamma=1.4$.  The most completely
resolved part is the soft XRB ($0.5-2~keV$), which is dominated by
unobscured AGN and QSOs but these sources have much steeper
(i.e. $\Gamma>1.4$) spectral shapes than the XRB and cannot explain
the spectrum at harder energies.  Therefore, a population of more
heavily obscured AGN, with flatter spectral shapes, is likely to be
involved in the production of the hard XRB.  The resolution of the XRB is less
complete at these higher energies, although {\em XMM-Newton\/} with
its sensitivity up to energies of $10~keV$, is making a major
contribution here.  The deepest X-ray surveys with {\em Chandra\/} and
{\em XMM-Newton\/} are beginning to reveal a population of highly
obscured AGN (eg. Hasinger et al. 2001). An obvious possibility is
that these highly obscured AGN, responsible for the hard XRB, are also
producing the CIRB.

The most direct way of testing this, of course, is to make X-ray
observations of the sources revealed by the deep SCUBA surveys.  The
studies which have so far been carried out (Almaini et al. 2001;
Fabian et al. 2000; Hornschemeier et al. 2000) suggest that SCUBA
sources are not generally X-ray sources.

In this paper we report our first results from {\em XMM-Newton\/}
X-ray surveys of the Canada France Redshift Survey fields (Lilly et
al. 1995).  The 3hr and 14hr fields coincide with large and deep
SCUBA surveys (Webb et al. 2002 \& Eales et al. 2000), and so we
report on the sub-mm/X-ray relation.  The 10hr field coincides with a
much smaller SCUBA survey, so it is not analysed in this paper.  It is
mentioned here for completeness, with full multi-wavelength analysis
of all three fields reserved for a future paper.

Assumed cosmology: $H_{0}=75~km~s^{-1}~Mpc^{-1}$,
$\Omega_{m}=1$ and $\Omega_{\Lambda}=0$.

\section{X-ray Data Reduction}
\label{data}
The data for the 3hr field were taken on 17th February 2001 by {\em
XMM-Newton\/} over a period of $51.5~ks$, using the thin filters and in 
imaging mode.  All three primary instruments gathered data (MOS 1, MOS
2 \& PN) as well as the optical monitor (OM) telescope.  This field is
centred on R.A. 03:02:38.60 Dec. +00:07:40.0. 

The 10hr field was similarly surveyed by XMM for $50.8~ks$, using the
thin filter for the PN instrument and the medium filter for the two
MOS instruments.  This field is centred on R.A 10:00:40.4
Dec. +25:14:20.0.

The 14hr field data was obtained from the public archive after the
proprietary period had expired, to complete our coverage of available XMM
data for the CFRS fields.  This data was first presented in Miyaji \&
Griffiths (2001).  Of the several available exposures of this
field, one was selected that most closely matched the exposures of the
other two fields.  The exposure was taken over $56.1~ks$, using thin
filters, and is centred on R.A. 14:17:12.0 Dec. +52:24:00.0.  We do
not discuss the Optical Monitor data here, for any of the above surveys.

The {\em XMM-Newton\/} raw data were processed using version 5.3 of an
ensemble of tasks collectively titled Science Analysis System
(SAS).  These tasks allow re-running of basic pipeline processes as
well as further data reduction tasks.  Initial pipeline processing of
the EPIC instruments is achieved through the running of the tasks
`emproc' for the two MOS instruments and `epproc' for the PN
instrument.  This produces the basic calibrated photon event files
(recording the time, position and energy of each photon) which are
used in all further processing.  These tasks also remove hot and
flickering pixels and columns.

Filtering of the event files is essential to obtain usable data, and
so this is the next step.  Non X-ray associated events such as cosmic
rays create patterns on the detectors that look different from the
impacting of X-rays.  These events are flagged and filtered out.  On
the other hand soft protons, produced by the sun and projected towards
Earth in solar flares, produce patterns that look identical to X-rays.
Therefore these need to be filtered out by generating a rate curve
displaying the count rate for an instrument as a function of time.
It is clear from this rate curve when flaring events occur during the 
observation, because of the vast increase in the count rate, and so
the affected time intervals can be removed from the data completely.
For the data in this work flaring events cause a loss of
$\sim20~per~cent$ of the total observing time in the 3hr field and
$\sim10~per~cent$ for the 10hr field.  The 14hr field was less affected by
flares and so only $\sim2~per~cent$ of data had to be removed. 

Specific energy channels can also be selected, enabling images in
different energy bands to be produced. Two bands are used in this
work, a soft band and a hard band, corresponding to $0.5-2~keV$ and
$2-10~keV$ respectively.  The low end $0.5~keV$ cutoff ensures
that X-ray emission from the Galaxy, which is greatest below this level, is
kept to a minimum.  Attenuation of soft X-rays by Galactic HI is also
more pronounced  at energies lower than this cutoff and so this too is
avoided.  The high energy limit of $10~keV$ is set by the instrument
response, which decreases rapidly at higher energies, more so for the
MOS instruments than the PN instrument.

The remaining noise is primarily due to the quiescent internal instrument
background caused by high energy particles interacting with the structure
surrounding the detectors.  There is also detector noise but this is
negligible and only becomes important at energies below $\sim0.5~keV$.
Due to the small number of photons involved in X-ray observations
additional noise is introduced by the poissonian statistics, and
ultimately this is the dominant source of error in the measurements.

Once filtering is complete images are produced for each instrument in
the two energy bands.  Free from bad pixels, flaring events and non
X-ray associated events the images can then be passed through the
source detection procedure.  This is a multi-stage process performed
on the combined bands for all three EPIC instruments simultaneously,
so that the maximum amount of information is used in determining the
source parameters.  The process is summarised here.

First, a sliding box source detection algorithm is run on the images
which flags any region that exceeds a minimum likelihood limit of 10
(equivalent to about $4~\sigma$) as a source.  The likelihood limit $L$
is defined such that $L=-\ln P$, where $P$ is the probability of
finding an excess above the background which is not due to a source.
The source list produced by this first pass is used to create a
``cheese'' map whereby all objects in the list are removed from the
image which is then interpolated using a spline to create a smooth
background ready for the second pass of the sliding box procedure.
The combined list from both passes is then fed into a maximum
likelihood (ML) algorithm which compares each object with a point
spread function (PSF) model to determine a confidence level for it being a
source.  All objects below a threshold (higher than for the sliding
box) are rejected leaving the final source list.  This method of
source detection is discussed in more detail in Valtchanov et al
(2001), and is compared against other methods for detecting sources in
{\em XMM-Newton\/} images.  In this work a ML limit of 15 is used, to
minimise the number of false detections.

In total 451 sources are detected in the three exposures, in one or
more instrument or band.  These are divided as follows: 146 in the 3hr 
field, 151 in the 10hr field and 154 in the 14hr field.  Full
discussion of the X-ray sources, together with catalogues, are reserved
for a future paper.  For this work only the areas observed in the Canada-UK
Submillimeter Survey (CUDSS) are considered.

Using two X-ray bands it is possible to define a quantity called the
hardness ratio, which gives an indication as to the basic spectrum of 
a source.  For this work it is defined as:

\[
 HR = {{N(H) - N(S)} \over {N(H) + N(S)}}
\]
where $N(H)$ and $N(S)$ are the counts observed for a source in the
hard and soft bands respectively, after correction for vignetting.
Higher values indicate a harder spectrum.

The minimum detected flux for these surveys is $\sim0.2\times
10^{-15}~erg~cm^{-2}~s^{-1}$ in the soft band, $\sim1\times
10^{-15}~erg~cm^{-2}~s^{-1}$ in the hard band and $\sim1\times
10^{-15}~erg~cm^{-2}~s^{-1}$ in the full ($0.5-10~keV$) band.
However, the cumulative source counts begin to deviate from the
expected power law at fluxes of about 0.6, 3 and $3\times
10^{-15}~erg~cm^{-2}~s^{-1}$ in the respective bands, thus we estimate 
that the surveys are $100~per~cent$ complete at these fluxes.

The astrometry for the source lists are compared to known QSOs in each 
field and corrected as appropriate.  The IDing process, to be
discussed more fully in a future paper, was carried out on the
corrected source positions.  As a second additional check for
systematic errors in the X-ray astrometry, the mean offset between the
X-ray sources and the IDs was calculated.  The residuals after
correction revealed no further systematic offsets.

XMM has a typical astrometric accuracy of $2\arcsec$ for sources near
the pointing axis, but this deteriorates gradually for progressively
larger off-axis angles.  At the depth of these surveys XMM does not
suffer from source confusion, which would become important for
exposure times of $\sim150-200~ks$.  The principle limit in this study
is the performance of SCUBA which has a beam size over twice that of
XMM.  In comparison {\em Chandra\/} has significantly better
resolution at $0.5\arcsec$ astrometric accuracy, and has a lower
detector background than XMM, but it is less sensitive (particularly
at energies $>8~keV$) and has a mirror effective area less than 1/5th
that of XMM.

\begin{figure*}
\begin{minipage}{14cm}
\psfig{file=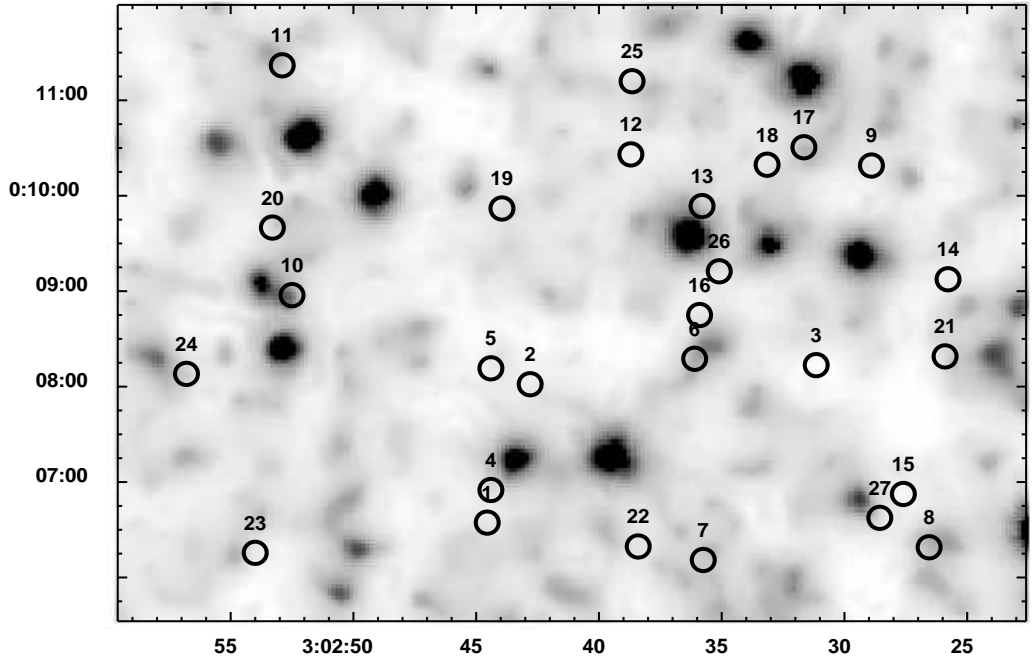,width=14.3cm,height=10cm}
\caption{\label{pic1} The central part of
the 3hr XMM survey showing the 3hr CUDSS region in the soft X-ray band,
for clarity.  Overlayed are $14\arcsec$ diameter circles at the
positions of the SCUBA sources. Numbers and positions are from Webb et
al. (2002).}  
\end{minipage}
\end{figure*}

\begin{figure*}
\begin{minipage}{14cm}
\psfig{file=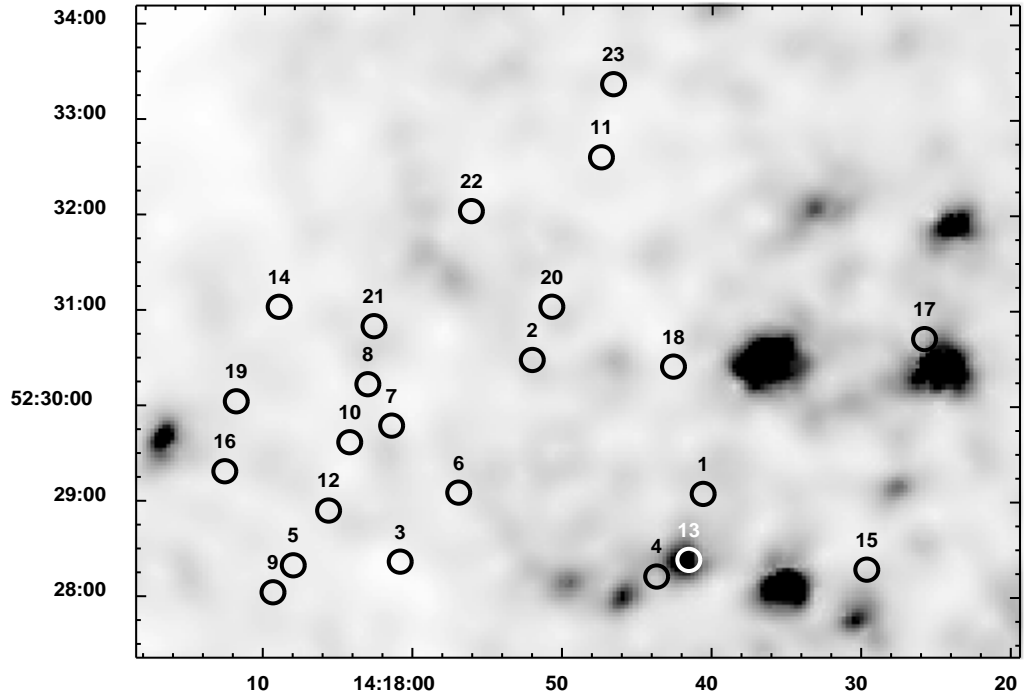,width=14.3cm,height=10cm}
\caption{\label{pic2} As fig~\ref{pic1} but for North-East part of the
14hr XMM survey showing the 14hr CUDSS region.  Numbers and positions
are from Eales et al. (2000) and Webb et al. (in preparation).  The
pattern of X-ray sources in this plot is mainly due to natural
variations in number density across the sky, and the domination of two 
QSOs near the right edge, although the far top left may suffer
slightly from the declining sensitivity of XMM towards the edge of its FoV.} 
\end{minipage}
\end{figure*}

The composite images shown in this work are a result of merging images 
from the PN and two MOS instruments together using the SAS task
`emosaic'.  This task is only used in this instance to produce a clear
image, and is not used for further analysis of the data, except for
the statistical tests described in section~\ref{coadd}.  See
figs.~\ref{pic1} and ~\ref{pic2}.

The majority of the work presented in this paper concentrates on the
3hr XMM field and the corresponding CUDSS region which lies in the centre
of the XMM FoV.  The 14hr field is also part of CUDSS but the survey
region lies towards the edge of the XMM FoV, and the X-ray data are
therefore less reliable. 

\section{SCUBA sources}
\label{scuba}
There are 27 sources detected at $850~\mu$m in the 3hr CUDSS field, spanning
a region of approximately $9\arcmin \times 6.4\arcmin$ (Webb et
al. 2002).  This lies in the central part of the {\em XMM-Newton\/}
FoV which is roughly circular with a diameter of $\sim30\arcmin$. 

The 14hr CUDSS field is approximately $7.7\arcmin \times 6.4\arcmin$
(Eales et al. 2000) and is centred about $8.5\arcmin$ North-East of
the XMM pointing axis.  This map contains 23 sources. 

\subsection{X-ray properties of the SCUBA sources}
\label{match}
The first thing to do when comparing sources in the same area but
different parts of the spectrum is to see if any of them match up.  An 
inexact match-up does not necessarily mean that two objects are
unassociated, because of the error in the positional accuracy of both
sets of objects, caused by the finite resolution of the instruments
involved.  For SCUBA the FWHM of the $850~\mu$m beam is
$\sim14\arcsec$, whereas for XMM the on-axis FWHM is $\sim6\arcsec$,
which increases with larger off-axis angles.  For the purposes of this 
work the XMM PSF can be considered constant, and equal to $6\arcsec$,
across the area containing the 3hr SCUBA sources, as this is a small
fraction of the total XMM FoV.  As for the 14hr field, the off-axis
angle of the CUDSS map means that the XMM PSF is not so well behaved,
and further analysis less reliable.  However, as an approximation the
FWHM has a median value of $\sim9\arcsec$ across the 14hr CUDSS map.

\subsubsection{The 3hr Field}
Of the 27 SCUBA sources in the 3hr field only one (CUDSS 3.10) is
possibly associated with a region of faint X-ray emission.  The flux of 
this region is below the flux limit of the survey, but it appears as a
faint patch in the smoothed image, fig.~\ref{pic1}.  This region was
picked up by an earlier version of the detection software at the
thresholds used, however after re-analysis with the updated software,
which has better calibration, this source is no longer detected using 
the same thresholds.  An additional degree of uncertainty exists for this
region, because it lies on a boundary between two PN CCD chips.
Source detection near chip edges is less precise than in the centre of
chips, so this possible associations should be regarded with extreme caution. 

Given the position of the SCUBA sources the probability that there is
a chance coincidence with an unrelated X-ray source within a distance
$r$ is given by poissonian statistics as: 

\[
 P = 1 - exp(-\pi n r^{2})
\]
where $n$ is the surface density of X-ray sources.

CUDSS 3.10 lies $\sim3.6\arcsec$ from the centre of the region of faint
X-ray emission (given by the detection with the earlier version of
SAS).  The surface density of X-ray sources at the flux limit of 
this survey is $n=(5.7 \pm 0.5) \times 10^{-5}$ per square arc second,
thus for this source, which is below the flux limit, $P>0.0023$.  A
deeper exposure of this region would determine if this X-ray region is
actually a source, and thus a significant association, or just noise. 

\subsubsection{The 14hr Field}
\label{14hr}
Figure~\ref{pic2} shows the CUDSS 14hr map as viewed in soft X-rays
with the SCUBA sources overlayed.  CUDSS 14.13 is clearly near a
significant X-ray source, and this is indeed detected by the XMM
source detection software.  This X-ray source has a soft band flux of
$(6.7\pm0.51)\times 10^{-15}~erg~cm^{-2}~s^{-1}$ and a hard band flux
of $(2.2\pm0.25)\times 10^{-14}~erg~cm^{-2}~s^{-1}$, assuming a power
law photon index of 1.4.  The positional offset is $4.8\arcsec$, and
the surface density of X-ray sources brighter than the flux of this
source is $n=(4.5\pm0.4) \times 10^{-5}$ per square arc sec.  This leads to
$P=(3.3\pm0.2)\times 10^{-3}$ for this coincidence. 

Webb et al. (in preparation) discuss this SCUBA source to some extent.  It
is identified with the optical source CFRS14.1157, which is
$1.2\arcsec$ from the position of our X-ray detection.  This source is 
also detected in the radio (Eales et al. 2000) and by ISO making it an
interesting source.  It also has a spectroscopically measured redshift 
of $z=1.15$ (Hammer et al. 1995).  The optical/NIR colours are very red,
$(I-K)_{AB}=2.6$, consistent with an irregular or spiral galaxy with
high extinction, and the X-ray hardness ratio is quite high ($\sim
-0.3$) implying that this is a fairly heavily obscured object.  HST
imaging of this object shows a disturbed morphology (Webb et al. in
preparation), suggesting the X-ray activity may be related to a possible
interaction.

By taking the measured redshift for this source, and the X-ray flux in 
the two bands, it is possible to estimate the column density of
neutral hydrogen responsible for the obscuration of the X-rays.  If we 
assume an intrinsic photon index of $\Gamma=2.0$ (Hasinger et
al. 2001) then a column density of $N_{H}=3.0 \times~10^{22}~cm^{-2}$
produces the correct X-ray fluxes. 

It is interesting to ask at this point whether, assuming an AGN is
responsible for the X-ray emission, it can also be responsible for the
sub-mm flux measured by SCUBA (through the heating of dust).  By assuming
that $3~per~cent$ of the bolometric luminosity of an unobscured AGN
(optical through to X-ray) is emitted in the $0.5-2~keV$ band (Page et
al. 2001), and correcting for the intrinsic absorption, the AGN has a total
luminosity of $(2.6 \pm 0.2)\times 10^{11} L_{\odot}$.

The far-IR luminosity on the other hand is $(5.1 \pm 1.5) \times 10^{12}
L_{\odot}$, assuming a single dust temperature of $40 K$ and
$\beta=1.5$.  This means that the AGN is 20 times less luminous and so
cannot possibly power the far-IR luminosity on its own.  The majority of
the far-IR luminosity must therefore be powered by star-formation, and 
in-fact Webb et al. (in preparation) estimate a star-formation rate of
$210~M_{\odot}~Yr^{-1}$ assuming that the far-IR luminosity is
produced in this way.  However, the far-IR luminosity is highly model
dependent and can be much lower, for example a single temperature dust
component of $<20 K$ gives a far-IR luminosity low enough to be equal
to the AGN luminosity.  Although this model is unlikely in light of the high
inferred star formation rate which would result in at least some warm
dust, and this would push up the luminosity greatly.  For example, a
two dust component model with 50 times as much cold dust ($15 K$) as
warm dust ($45 K$) and $\beta=2$ (e.g. Dunne \& Eales 2001) has a
far-IR luminosity of $(2.2 \pm 0.7) \times 10^{12} L_{\odot}$, making
it 8 times as luminous as the AGN.  Thus, the conclusion that this
source is dominated by star-formation and not an AGN is hard to avoid. 

Further evidence for the star-formation activity of AGN host galaxies
can be found in the $8~mJy$ SCUBA survey fields (Ivison et al. 2002).
The X-ray detected SCUBA sources in these fields ($15~per~cent$) are
consistent with obscured AGN, but the AGN bolometric luminosities are
not sufficient to power the far-IR luminosities, unless the X-ray
emission is attenuated by Compton thick ($N_{H}> 10^{24}~cm^{-2}$)
material.  If not, star-formation is likely to be responsible instead.

Page et al. (2001) draw a similar conclusion for high redshift X-ray
selected sources that have a SCUBA detection.  For their sample the
AGN luminosity and far-IR luminosity are more closely matched, being
at most a factor of 4 different in favour of the far-IR luminosity.
The main differences in this case are that the AGN in their sample are at
higher redshift ($z=1.5 - 3$) and are more luminous ($L_{AGN} > 4.36
\times 10^{12} L_{\odot}$) than CUDSS 14.13.  The HI column
densities in their sample are similar to CUDSS 14.13.

Whereas Page et al. (2001) target X-ray sources with SCUBA, a reverse
study whereby bright SCUBA sources are observed with the {\em
Chandra\/} X-ray observatory was carried out by Bautz et
al. (2000).  They measure the AGN/far-IR luminosity for two bright, 
lensed sub-mm sources at high redshift and find that the AGN is 
responsible for the majority of the far-IR luminosity in one and
$\sim~40~per~cent$ in the other, implying that some other power source
must be responsible for the deficits.

\subsection{Statistical analysis}
\label{coadd}
The fact that only one out of 50 SCUBA sources matches an X-ray source
does not mean that the population as a whole is not significantly
emitting X-rays.  A simple way to test this is to coadd the X-ray
counts associated with each SCUBA source and see if the average X-ray
flux of the SCUBA sources is statistically significant.  We refer to
this as the coadding technique.

The CUDSS 3hr map is located in the centre of the XMM FoV which lies
entirely in the central CCDs of the two MOS arrays.  However, the PN
array contains many chip boundaries in this region, and several SCUBA
sources lie on or near one of these boundaries.  Therefore, to avoid
possible problems caused by inconsistencies between sources, the PN
data is not used in the analysis.

Due to the 14hr CUDSS map being off-axis in the XMM survey we consider
the 3hr field to be more useful and accurate for study.  However, for
consistency, the 14hr field is analysed in the same way as the 
3hr map, but is considered separately because of the difficulty in
combining the results from the two regions in any sensible way.  The
following description relates to the 3hr field, with differences for
the 14hr field noted where relevant.

\begin{figure*}
 \subfigure[\label{hist1}]{\psfig{file=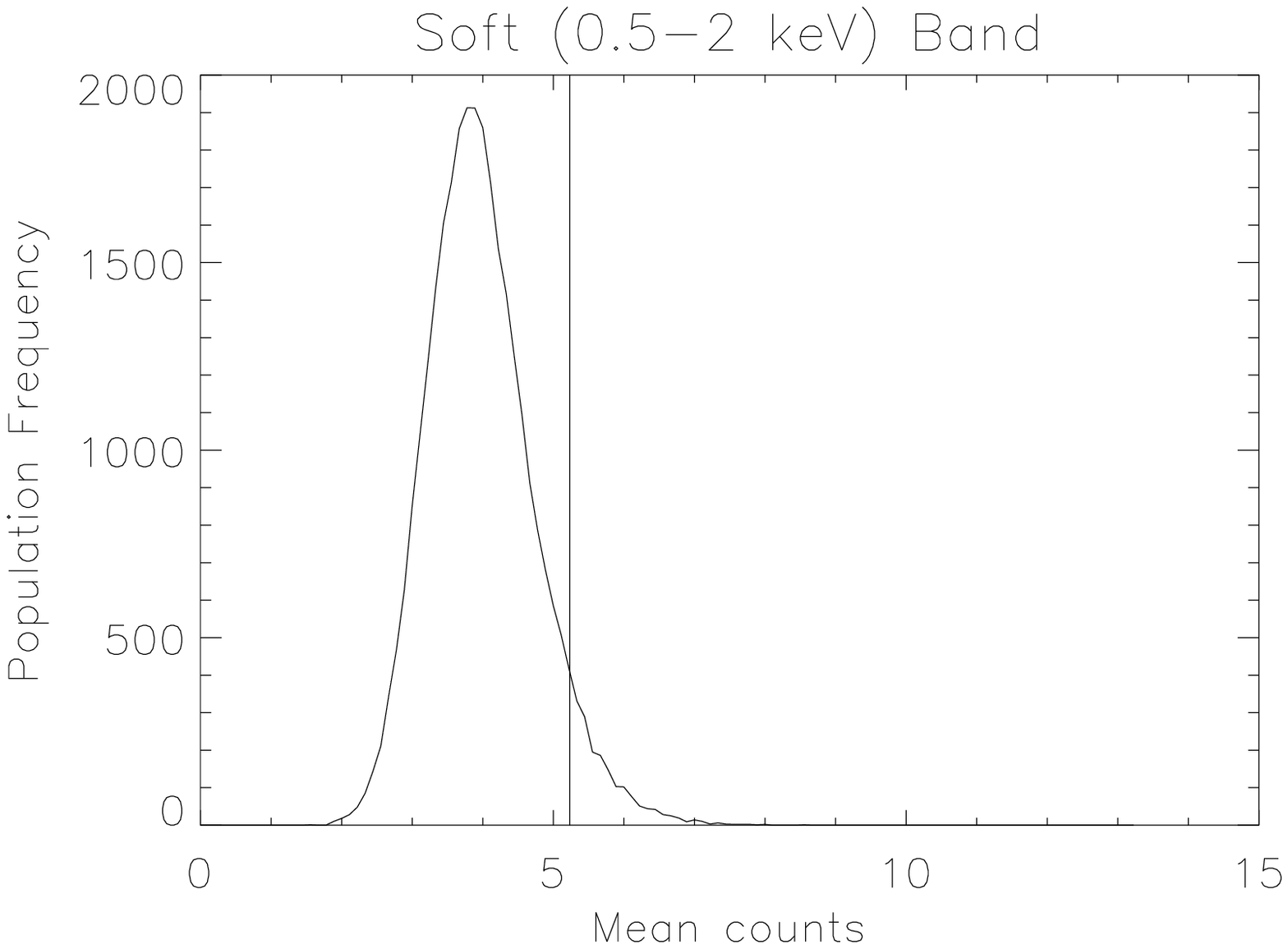,width=8cm,height=6cm}}
 \subfigure[\label{hist2}]{\psfig{file=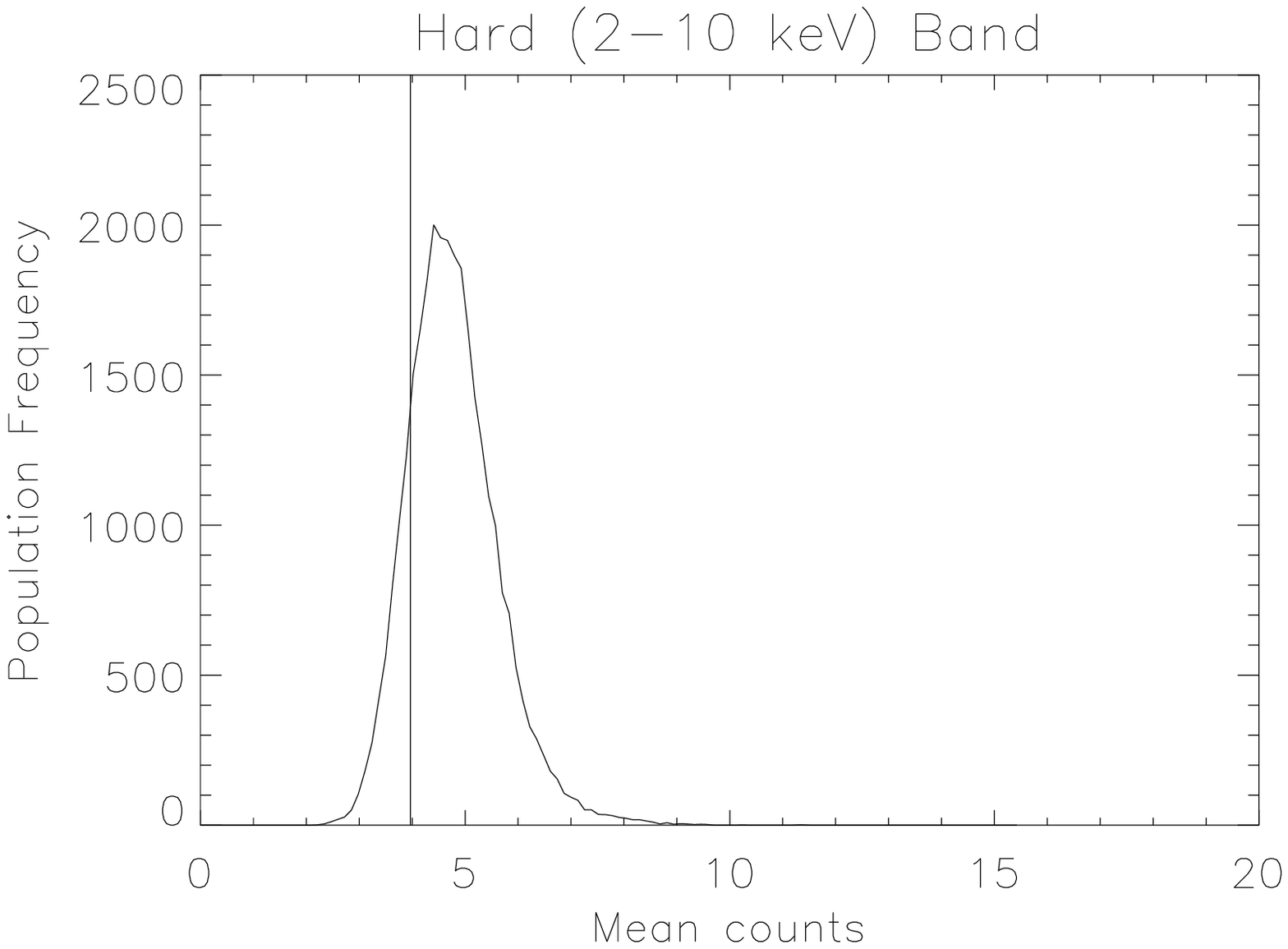,width=8cm,height=6cm}}
 \caption{Histograms showing the distribution of mean X-ray counts
associated with each source for 30,000 artificial SCUBA samples in the 
3hr field.  Vertical lines represent the mean counts for the real
SCUBA sources.  See text for details.} 
\end{figure*} 

\begin{figure*}
 \subfigure[\label{hist3}]{\psfig{file=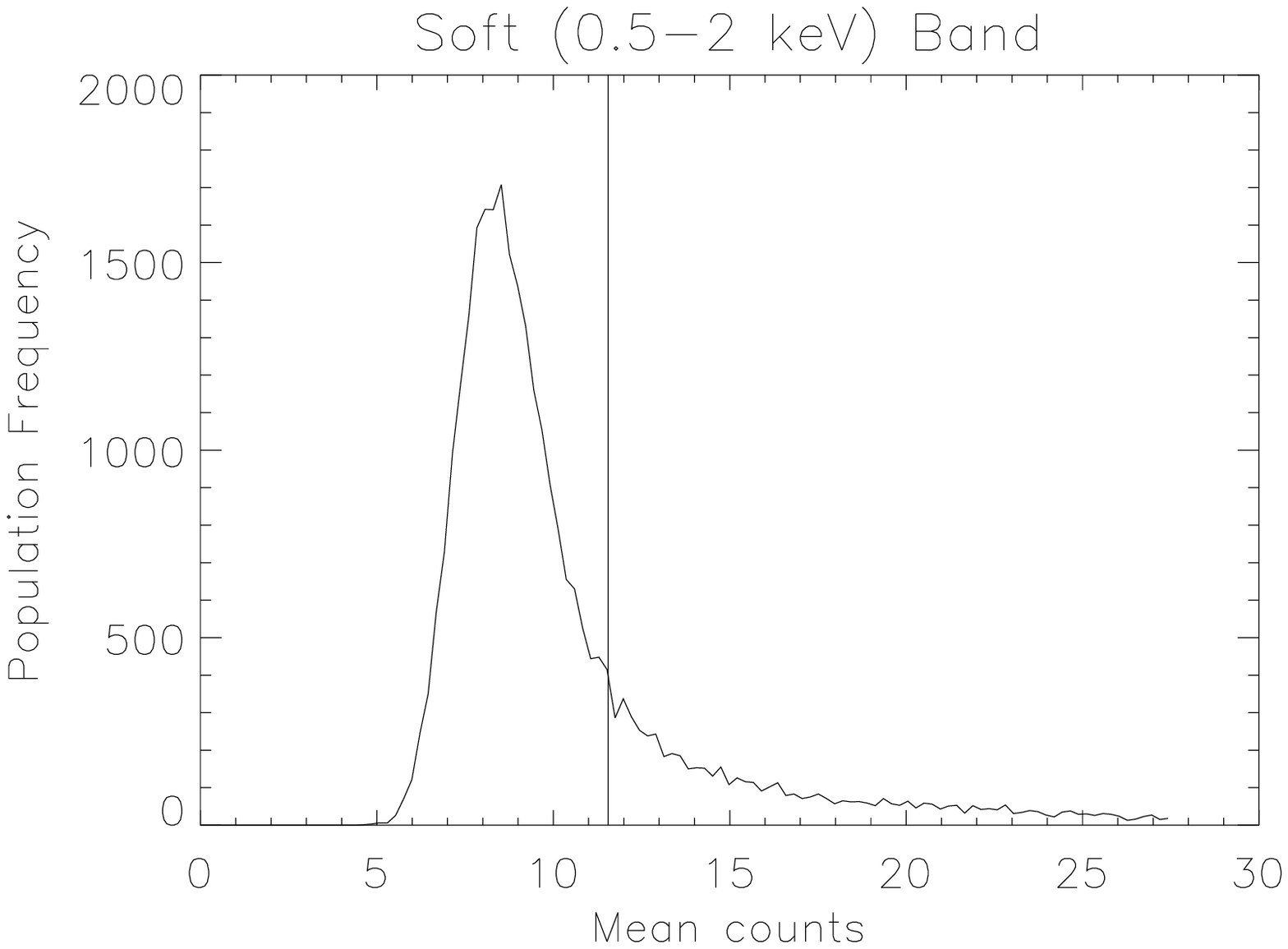,width=8cm,height=6cm}}
 \subfigure[\label{hist4}]{\psfig{file=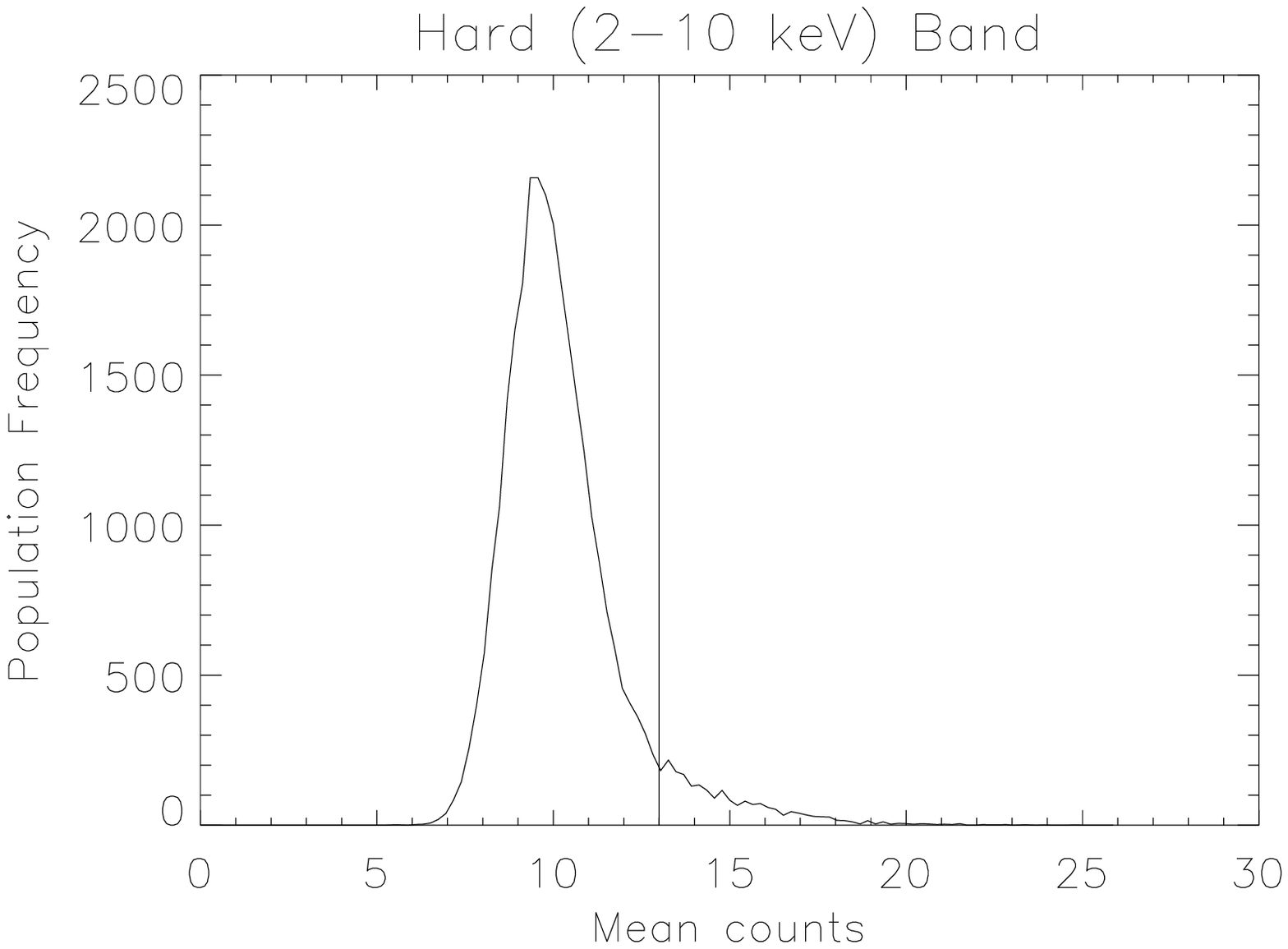,width=8cm,height=6cm}}
 \caption{As for figures~\ref{hist1}~\&~\ref{hist2} but for the 14hr
field.  This plot highlights the higher background in this field
compared to the 3hr field, a feature of the survey itself rather than
the poorer response of the instrument for large off-axis angles.}
\end{figure*}

The XMM optics spread out photons from a point source into several
pixels in the images (the PSF).  Although the pixel at the co-ordinates
of the source should contain the peak of the emission, neighbouring pixels
also contain information from the source.  This effect is important for
faint X-ray sources in particular because in some cases there may be
no actual photon counts in the pixel corresponding to position of the
source.  Therefore, for analysis of the SCUBA sources in the X-ray
images, the information from the whole PSF of XMM needs to be
recovered and incorporated into the central pixel. 
  
Images were accumulated in each band for the two MOS instruments and
superimposed, as described at the end of section~\ref{data}.  They
were then convolved with a 2D Gaussian of $FWHM=6\arcsec$, the FWHM of 
the on-axis XMM beam ($9\arcsec$ for the 14hr field to reflect the
larger, off-axis, PSF).  This creates a map in which the signal in each
pixel is the best possible estimate of the X-ray signal at that
point.  By smoothing the image with the XMM PSF (or a reasonable model
for it in this case), the information in the central pixel contains
all the information from the surrounding pixels, weighted according to
how likely it is that the photons detected in those pixels came from the
source.  This technique for making the best estimate of the flux at a
given point has been widely used in other wavebands (eg. Phillipps \&
Davies 1991; Eales et al. 2000), and is similar to the stacking
technique used in the analysis of the {\em Chandra} Deep Fields
(eg. Brandt et al. 2001).

The mean X-ray flux of the 27 SCUBA sources was determined for each
energy band.  To estimate the significance of the values a Monte-Carlo 
simulation was used.  30,000 random samples of SCUBA sources were
produced at randomly selected positions within the CUDSS region, each
with 27 sources.  The mean X-ray flux for each sample was measured in
the same way as for the real sample.  The distribution of these means
is shown in figs~\ref{hist1} and~\ref{hist2}.  The number of trial
samples for which the mean equals or exceeds the observed sample mean
is then calculated (See figures~\ref{hist3} and~\ref{hist4} for the 23
SCUBA sources in the 14hr field).  A low number of trials exceeding
the observation indicates a significantly important measurement. 

Out of 30,000 random trials 2,010 equalled or exceeded the observation
($\sim6.7~per~cent$) for the soft band and 25,800
($\sim86.0~per~cent$) for the hard band (see  fig.~\ref{hist1}
and~\ref{hist2}).  A more restricted sample was also tested, in case
some of the less secure SCUBA detections were in fact false.  The 9
SCUBA sources with the highest flux produced $\sim63.3~per~cent$ and
$\sim62.4~per~cent$ of trials that equalled or exceeded the
observation, for the soft and hard bands respectively.  Thus the SCUBA
population is not detected, and is lost within the unresolved X-ray
background.  Converting the mean counts per SCUBA source into a mean
flux gives a $99~per~cent$ upper limit estimate of $1.25\times
10^{-16}~erg~cm^{-2}~s^{-1}$ in the soft band and $6.62\times
10^{-16}~erg~cm^{-2}~s^{-1}$ in the hard band, the true values falling 
below these limits $99~per~cent$ of the time.

As an additional test, to take the accuracy of the SCUBA positions into 
account, we repeated this test by selecting the brightest X-ray pixel
within a search radius of $4\arcsec$ of the given SCUBA position, instead of
just taking the pixel value at that position.  No significant
difference was obtained, as the mean fluxes for both the randomly
generated samples and the true samples increased by a similar number
of X-ray counts, leading to similar significances and mean X-ray flux limits.

For the 14hr field $22.7~per~cent$ and $7.4~per~cent$ of trials
equalled or exceeded the observation, in the soft and hard bands
respectively (see figures~\ref{hist3} and~\ref{hist4}).  The flux
limits are less stringent than the 3hr field at $8.0\times
10^{-16}~erg~cm^{-2}~s^{-1}$ in the soft band and $1.3\times
10^{-15}~erg~cm^{-2}~s^{-1}$ in the hard band.  This is likely to be a 
consequence of the 14hr CUDSS region being off-axis in the XMM survey.
The vignetting of the telescope increases the flux limit towards 
the edge of the FoV because of the reduced effective exposure time
compared to on-axis.

\section{sub-mm properties of X-ray sources}
\label{xray}
Taking the reverse approach, the sub-mm properties of the 18(16) X-ray
sources within the 3hr(14hr) SCUBA maps can be determined.  The SCUBA maps are
heavily confused, not only from the positive sources but also the
negative side lobes produced by the chopping procedure.  Therefore, the
SCUBA sources not associated with X-ray sources are removed from the
map, including the side lobes.  This reduces the confusion from known
sources, in order to better test the low level emission in the map
that is unresolved but may still be real, and associated with X-ray
sources.  After this procedure, the weighted mean $850~\mu$m flux of
the 3hr X-ray population is found to be $0.48\pm0.27~mJy$.  The 14hr
field yields a mean $850~\mu$m flux of $0.35\pm0.28~mJy$.  These are
not significant detections, but are tentatively suggestive of dust
emission.  

These measurements are in contrast to the mean sub-mm flux of
$1.69\pm0.27~mJy$ obtained by Barger et al. (2001) for a sample of 136
X-ray sources selected in the $2-8~keV$ band, detected in the Chandra
Deep Field North.  Although for a restricted sample of soft X-ray
sources with $\Gamma>1$ they find the mean sub-mm flux is much lower
at $0.89\pm0.24~mJy$, consistent with our $3~\sigma$ upper limits of
$0.81$ and $0.84~mJy$ for the two fields.  This is also consistent
with Almaini et al. (2001) who measure a noise weighted mean of
$0.89\pm0.3~mJy$ for their X-ray sources.  It is perhaps not
surprising that Barger et al. (2001) find many X-ray sources with hard
spectra ($\Gamma<1$), for which they measure a mean sub-mm flux of
$1.77\pm0.21~mJy$, since their sources are selected in a hard X-ray
band.  In contrast our sources are selected in the soft $0.5-2~keV$
band as well as the hard $2-10~keV$ band, and so there are many more
sources with soft rather than hard spectra in our surveys.  These soft 
sources dominate our sub-mm measurement which could partially explain
our lower values.

An alternative explanation for the differing mean sub-mm measurements
in different studies may come from the different methods used in
calculating them.  Not removing SCUBA sources that do not coincide
with X-ray sources will result in higher residual sub-mm measurements
than our method.

The mean soft X-ray flux for the X-ray sources in these
two regions are $(1.8\pm0.1)\times 10^{-15}~erg~cm^{-2}~s^{-1}$ for the 3hr
CUDSS map and $(7.8\pm0.4)\times 10^{-15}~erg~cm^{-2}~s^{-1}$ for the 14hr
CUDSS map, measured in the soft band.

\section{Discussion}
\subsection{AGN Verses Star-formation}
The lack of X-ray/SCUBA coincidences suggests three main
possibilities.  Either the X-ray survey is not deep enough to detect
the AGN that may exist within the SCUBA sources; the AGN are obscured
by Compton thick material, leading to very little nuclear radiation
escaping unhindered ($N_{H}>10^{24-25}~cm^{-2}$); or the SCUBA sources
do not contain AGN.

\begin{figure*}
 \subfigure[\label{plot1}]{\psfig{file=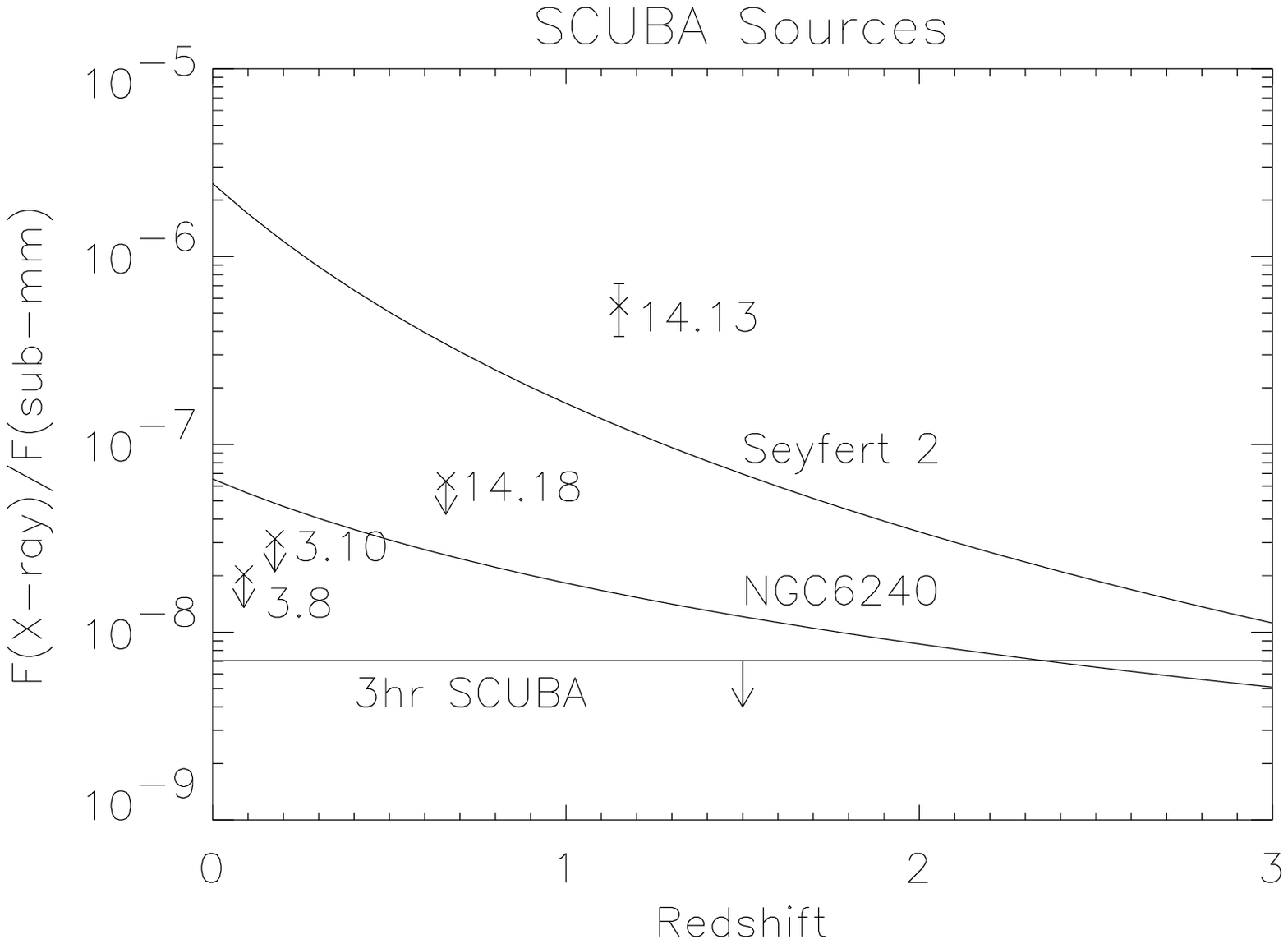,width=8cm,height=6cm}}
 \subfigure[\label{plot2}]{\psfig{file=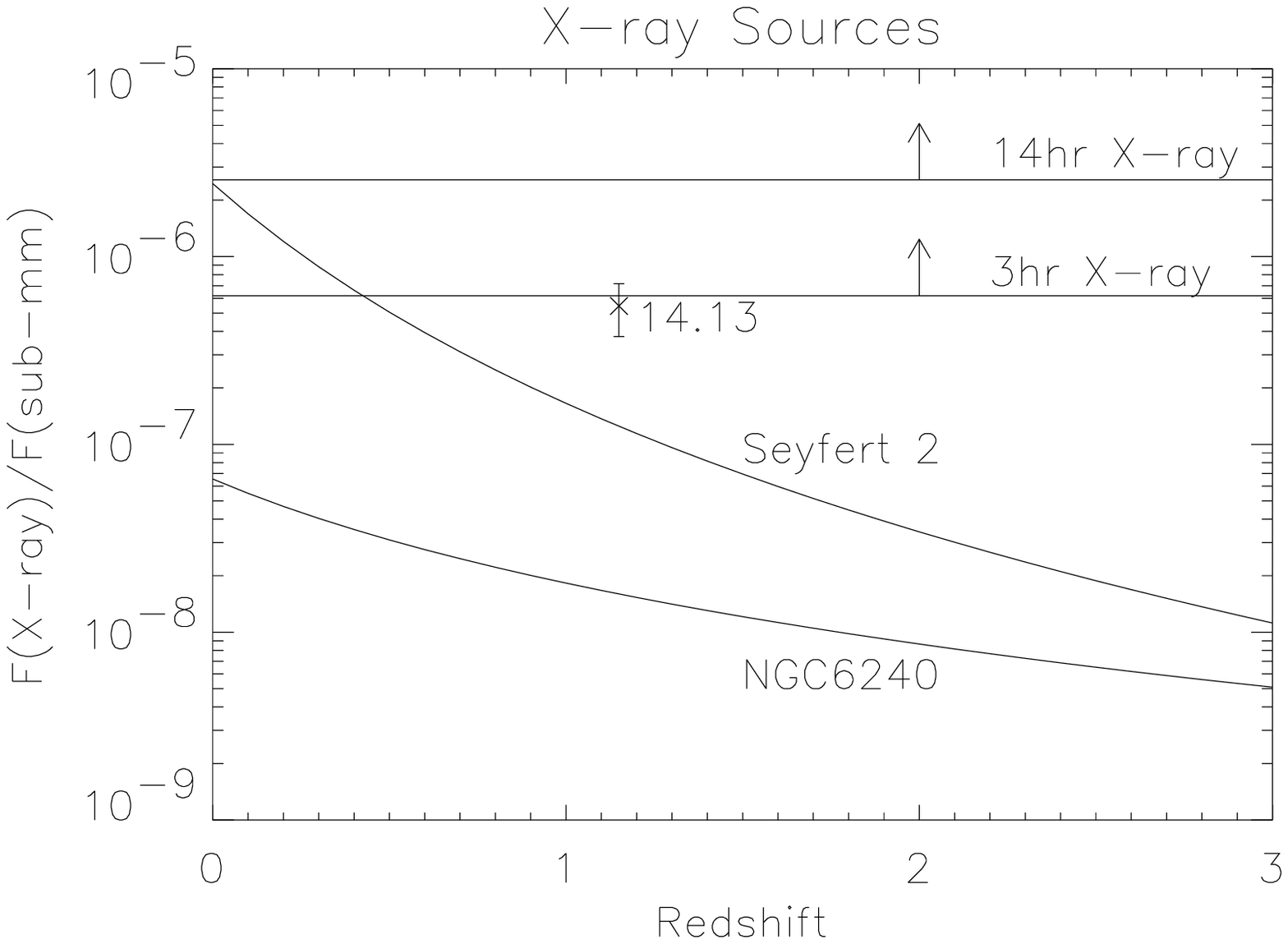,width=8cm,height=6cm}}
 \caption{\label{plot}X-ray to sub-mm flux ratios for the SCUBA sources
(\ref{plot1}) and the X-ray sources (\ref{plot2}).  Template SEDs of
NGC6240 and a Seyfert 2 Galaxy are plotted for comparison.  Horizontal
lines represent the ratios of the measured mean fluxes for the 3hr
sub-mm sample (\ref{plot1}), and the X-ray samples (\ref{plot2}).
The only definite sub-mm/X-ray coincidence is plotted in
\ref{plot2}, and is from Eales et al. (2000) with redshift from
Webb et al (in preparation).  The two 3hr SCUBA sources with
secure IDs and spectroscopic redshifts (Webb et al. 2002) are
plotted in \ref{plot1} along with the only other 14hr SCUBA source
with a secure ID and redshift (Webb et al. in preparation).  The low z 
3hr sources are not representative of the SCUBA population as a whole, 
which in general lie at higher z, and so do not necessarily have the
same properties.  Their optical counterparts both show merger
morphology.  Source labels are the CUDSS reference numbers.}
\end{figure*} 

Redshift information on the X-ray and SCUBA populations lends support
to the separate population hypothesis.  Recent deep X-ray surveys
(Hasinger et al. 2001; Rosati et al. 2002; Mainieri et al. 2002) have
determined the redshift distribution for large fractions of their
sources.  With a median redshift of $<1$ these surveys differ from the
AGN synthesis models (Gilli, Salvati \& Hasinger 2001) which predict
higher peak redshifts 0f 1.3-2.  In contrast the deep SCUBA
populations have been placed at much higher redshifts, typically
$z\sim2$ or greater (Ivison et al. 2002; Chapman et al. 2003), which
would naturally explain the small overlap between the two observed
populations. 

Can we be sure that SCUBA sources are powered by star-formation and
not by AGN?  The latter is still a possibility, despite the low SCUBA/X-ray
coincidence, as the AGN may be heavily obscured and as such not
visible to this X-ray survey.  NGC6240, for example, is known to be a starburst
galaxy that also contains a heavily obscured AGN ($N_{H}>2\times
10^{24}~cm^{-2}$) (eg. Iwasawa \& Comastri 1998; Lira et al. 2002).
To compare the SCUBA sources to NGC6240 we use the SED for a
`high-reddening' starburst galaxy taken from Schmidt et al (1997).
This fits the measurements for NGC6420 well (Lira et al. 2002).  We
use this SED as a basic template, modified for an X-ray photon index
of 1.4 to approximate the broad characteristics of NGC6240 (Iwasawa \&
Comastri. 1998).  We calculate the ratio between the soft X-ray flux
and the $850~\mu$m flux in the observed frame as a function of
redshift, and this is plotted in fig.~\ref{plot}.  We also plot the
X-ray/sub-mm flux ratio for a typical Seyfert 2 galaxy with respect to
redshift, using a template also taken from Schmidt et al. (1997).

The upper limit on the ratio of the mean X-ray to sub-mm flux for the SCUBA
sample is consistent with the template for NGC6240 as long as the
redshifts of the SCUBA sources are $>2.3$.  This is consistent with
redshifts measured for many SCUBA sources (Ivison et al. 2002; Chapman
et al. 2003), and so we conclude that in general it is still possible
that most SCUBA sources may contain Compton thick AGN like NGC6240.
Almaini et al. (2001) rule out the possibility that SCUBA sources are
QSOs, unless they are Compton thick and at very high redshift, and
show that they are consistent with a starburst template at $z>2$. This
is in general agreement with our results.  The SCUBA sources with
secure identifications and spectroscopic redshifts (Webb et al. 2002;
Webb et al. in preparation) are also plotted for the two fields.

A similar plot for the X-ray sources is shown in fig.~\ref{plot2}.
Note that the 14hr field has a higher mean X-ray flux because this field
is dominated by several bright X-ray sources (two are known QSOs,
Schade et al. 1996). 

In an alternative approach we can calculate the far-IR and AGN
luminosities of the ensemble of X-ray sources in the 3hr CUDSS map, as
we did with CUDSS 14.13 (section~\ref{14hr}).  We use the mean SCUBA
flux measurement of $0.48~mJy$ at $850~\mu$m, and assume a column
density of $10^{22}~cm^{-2}$ to be representative of the X-ray
sources.  For any reasonable far-IR SED the AGN luminosity, from
optical through to X-ray, never exceeds the far-IR luminosity for
redshifts below 2.  If we require the entire sub-mm flux
to be produced by the AGN then the X-ray sources must either be
modestly absorbed at very high redshifts, or extremely highly
absorbed and at lower redshifts.  This shows that in general the
bolometric luminosities of the X-ray sources in this field are
probably dominated by star-formation in the same way as CUDSS 14.13
is.  All of this, however, is highly speculative because of the highly
model dependent nature of the far-IR luminosity calculation, and the
marginal (not even $2~\sigma$) sub-mm detection. 

\subsection{Extra-Galactic Background Radiation}
The following analysis is based on the 3hr CUDSS region.

The upper limit on the average X-ray flux of the SCUBA sample also
allows us to place an upper limit on the contribution the SCUBA
population makes to the XRB.  Our sub-mm sample constitutes about
$20~per~cent$ of the extra-galactic background at $850~\mu$m, and so we scale 
our upper limit by a factor of 5 to calculate an upper limit on the
contribution dust sources make to the XRB (assuming that the
X-ray/sub-mm ratio is not dependent on sub-mm flux).  The sample provides a
maximum of $3.3~per~cent$ of the XRB at $0.5-2~keV$, and so the
population as a whole must contribute no more than $16.5~per~cent$ in this
band.  In the hard band the SCUBA sample contributes an upper limit of
$6.1~per~cent$ to the XRB and so the population as a whole provides at most
$30.7~per~cent$.  It is clear that sub-mm sources do not dominate the
X-Ray background at low energies, but it remains to be seen if their
contribution to the peak of the XRB ($\sim30~keV$) is more
significant.  

Taking the opposite approach we can estimate the contribution of AGN to
the $850~\mu$m background.  The simplest way is to convert our
$3~\sigma$ upper limit sub-mm flux, as measured in section~\ref{xray},
of the detected X-ray sources into a smooth background by multiplying
it by the number density of X-ray sources detected by our
survey. Comparing this to the intensity of the CIRB at $850~\mu$m
gives an estimate of a $2.3~per~cent$ contribution from AGN.  Assuming
the 3hr field is a typical extra galactic region and that the X-ray
sources are indeed AGN then this is a little low compared with
theoretical models (eg. Almaini, Lawrence \& Boyle 1999).  Our X-ray survey,
however, only resolves about $32~per~cent$ of the X-ray background in
the soft band.  If we assume that the background we have not resolved
has the same X-ray/sub-mm ratio as our X-ray sample, then
$7.2~per~cent$ of the sub-mm background is produced by the sources
making up the X-ray background.  This may be too low because the
fainter X-ray sources may be more heavily obscured, and the results
from the {\em Chandra\/} Deep Fields do suggest that the ratio
decreases at lower X-ray flux (Barger et al. 2001).  However, we have
argued above that much of the sub-mm emission from X-ray sources is
from dust heated by star-formation and not by AGN, and thus the
estimate above is really an upper limit on the contribution of AGN to
the sub-mm background.  These two arguments work in opposite
directions and may act to cancel each other out.  

The contribution of the SCUBA sources to the X-ray background and of
the X-ray sources to the sub-mm background strongly suggest the two
backgrounds are disjoint, with the sub-mm background mostly produced
by stellar nucleosynthesis and the X-ray background by accretion on to
black-holes. 

\section{Concluding Remarks}
This study adds weight to the growing body of evidence pointing
to the fact that SCUBA sources may indeed contain active nuclei,
but that their presence is secondary to starburst activity as a power
source for their high far-IR luminosities.  In some cases it is possible
to view the active nucleus directly using current X-ray satellites,
but there are still many SCUBA sources that show no evidence of X-ray
emission.  Higher and higher column densities of HI progressively wipe
out X-ray emission up to higher energies, and as such are capable of
putting AGN beyond detectability by current X-ray detectors sensitive only up
to $\sim10~keV$.  If this is the case for the majority of SCUBA
sources then the scenario outlined above will need to be modified, as
it is possible for the entire sub-mm luminosity of a SCUBA source to
be powered by a very powerful but highly obscured AGN, with no need
for a starburst.  How common such highly obscured systems are is
likely to remain uncertain until X-ray instruments become available
with good spatial resolution and good sensitivity above $10~keV$, in
order to detect obscured AGN beyond $z=2$, where many SCUBA sources are
being identified.  The deepest current X-ray surveys however (Alexander et
al. 2003), suggest that the low detection rate of X-ray emission from
SCUBA sources in less deep surveys is more likely to be due to low AGN
luminosity rather than heavy obscuration, which would tend to support
the statement at the start of this section.

Star-formation may also be the dominant source of the far-IR bolometric
luminosity of galaxies containing relatively bright X-ray emitting
AGN.  Certainly the significant sub-mm measurements of X-ray sources
in other studies indicates that this is the case.  Even a modest
sub-mm flux equates to a high far-IR luminosity, whatever dust SED is
assumed.
 
We conclude that in general the two extra-galactic backgrounds are
mainly produced by different processes, with the sub-mm background
being predominantly produced by dust being heated by starlight, and
the X-ray background being dominated by accretion onto super-massive
black-holes. 

\section*{acknowledgements}
TJW wishes to thank Mat Page for his invaluable advice about the XMM
data reduction, Loretta Dunne for her help with the far-IR luminosity
calculations and the referee for making useful comments leading to a
clearer paper.  TJW acknowledges the support of a departmental
postgraduate grant.  SAE thanks the Leverhume Trust for a research
fellowship.  TXT acknowledges the partial financial support of NASA
grant NAG5-9900.

\end{document}